\documentclass[conference]{IEEEtran}
\IEEEoverridecommandlockouts
\usepackage{cite}
\usepackage{amsmath,amssymb,amsfonts}
\usepackage{algorithmic}
\usepackage{graphicx}
\usepackage{textcomp}
\usepackage{xcolor}
\def\BibTeX{{\rm B\kern-.05em{\sc i\kern-.025em b}\kern-.08em
    T\kern-.1667em\lower.7ex\hbox{E}\kern-.125emX}}
\begin{document}

\title{Plausible deniability for privacy-preserving data synthesis\\
}

\author{\IEEEauthorblockN{1\textsuperscript{st} Song Mei}
\IEEEauthorblockA{\textit{Huazhong University of Science and Technology} \\
Wuhan, China \\
meisong@mail.hust.edu.cn}
\and
\IEEEauthorblockN{2\textsuperscript{nd} Zhiqiang Ye}
\IEEEauthorblockA{\textit{Wuhan Siwei Tongfei Network Technology Company Limited} \\
Wuhan, China \\
Zhiqiang@mail.hust.edu.cn}

}

\maketitle

\begin{abstract}
In the field of privacy protection, publishing complete data (especially high-dimensional data sets) is one of the most challenging problems. The common encryption technology can not deal with the attacker to take 
differential attack to obtain sensitive information, while the existing differential privacy protection algorithm model takes a long time for high-dimensional calculation and needs to add noise to reduce data accuracy, which is not suitable for high-dimensional large data sets. In view of this situation, this paper designs a complete data synthesis scheme to protect data privacy around the concept of "plausible denial". Firstly, the paper provides the theoretical support for the difference between "plausible data" and "plausible data". In the process of scheme designing, this paper decomposes the scheme design into construction data synthesis module and privacy test module, then designs algorithm models for them respectively and realizes the function of privacy protection. 
When evaluating the feasibility of the scheme, the paper selects the Results of the 2013 community census in the United States as the high-dimensional data set, uses the simulation program that is based on Python to test and
analyzes the efficiency and reliability of the data synthesis scheme. This portion focuses on the evaluation of
the privacy protection effectiveness of the scheme. In order to text this part, the paper tests the success rate of the synthetic data passing the differential privacy test, and uses LR (Logistic Regression) classifier and 
SVM(Support Vector Machine)classifier to verify the "indistinguishable" characteristics of the synthetic data. Finally, the article summarizes the full text and puts forward the direction that can be improved based on the existing model.  
\end{abstract}

\begin{IEEEkeywords}
data synthesis, privacy protection, differential privacy, classifier
\end{IEEEkeywords}

\section{Introduction}
The era of big data has arrived, and the issue of ensuring the security of data sets has become a strategic issue that directly affects the stability and security of the country and society. Data sets containing massive amounts of information are already recognized as valuable assets, and their analysis and utilization will create enormous added value. Therefore, individuals, organizations and even countries are paying more and more attention to the importance of data security. Driven by the huge value hidden in big data, many enterprises or organizations collect, process, use and publish personal information without restriction. Not only that, different departments of the same company and different companies reach an agreement to share user information. This has brought serious data privacy problems, that is, data privacy leakage, which in turn leads to the leakage of users' personal privacy information, such as users' assets, liabilities, health, marriage, political beliefs, etc.\cite{1}\cite{2}

Substituting a synthetic dataset for a user's real dataset is a common approach to preserve data privacy. The rapid development of the Internet has led to the rapid development of information technology and a large increase in information exchange. At present, compared with the explosive development of big data, data privacy protection is not perfect. In order to ensure the security of data privacy, it is necessary to find a safe and effective privacy protection scheme to deal with this huge challenge. Substituting a synthetic dataset for a user's real dataset is a common approach to preserve data privacy. This paper intends to combine the concept of "plausible denial" and differential privacy technology to design a data synthesis scheme to protect data privacy.\cite{3}

At present, some researchers have noticed the data security issues caused by the release of datasets, and have adopted differential privacy mechanisms to deal with interactive counting queries of databases. These research works can solve the problem of information leakage of sensitive data in interactive queries, and meet the needs of privacy protection to a certain extent. However, such schemes need to be implemented based on strict differential privacy mechanisms. Taking the Laplace mechanism as an example, its typical method is to add random noise to the data generation model. In non-interactive query, there are unfavorable factors such as excessive computational overhead on high-dimensional data sets. will introduce too much noise to deal with the problem of publishing a complete high-dimensional data record set.\cite{4}

Inspired by the works of Bindschaedler and others, this paper focuses on the concept of plausible denial and promotes it as a feasible data protection scheme, designing two modules of data generation and privacy testing. The data generation module adopts statistical principles to generate synthetic data according to its statistical distribution; each piece of synthetic data needs to pass the privacy test module, and if it passes the test, it will be released; otherwise, the data will be discarded, so as to realize the privacy protection of publishing a complete high-dimensional data record set utility. By separating the data generation and privacy testing modules, the scheme can improve the synthesis efficiency, generate synthetic data more efficiently, and facilitate large-scale use; it is universal and fills in the shortcomings of differential privacy based on the index mechanism that is difficult to deal with non-interactive queries. The data generation model does not need output disturbance (artificially added random noise), which saves computing overhead; this scheme will discard the synthetic data that has not passed the privacy test, and by adjusting the parameters of the privacy test module, it is possible to publish high-dimensional data with high reliability of privacy protection. data set.
\cite{5}\cite{6}

After completing the design of the above theoretical model, use cmake assembly software to complete the code compilation of the program, and use the data of the 2013 American Community Survey to verify, analyze the synthetic data efficiency and data protection utility of plausible denial of privacy protection, and verify the program feasibility.

The first chapter of this paper is the introduction part, which is used to explain the research background of the privacy protection scheme. The second chapter is used to describe the basic knowledge used in the article. The third chapter introduces the implementation process of the data synthesis scheme. The fourth chapter simulates and verifies the synthesis efficiency of the data synthesizer, analyzes the usefulness of the synthesized data, and focuses on verifying the privacy protection utility of the synthesized data. Finally, summarize the full text.\cite{7}\cite{8}

\section{Related Basic Knowledge}

\subsection{The concept of "specious denial"}
Paradoxical denial is understood as: an attacker with any knowledge background cannot infer a certain data record from the published data set, and the data has a greater impact on the statistical results than other data. That is, for a certain privacy parameter k, when k>0, at least k pieces of data can be observed, which have the same impact on the final output statistics, and the mechanism is said to satisfy plausible denial.

Compared with common differential privacy mechanisms such as the Laplacian mechanism, the plausible denial mechanism does not need to add artificial noise to the generated data, but divides the process of synthesizing privacy-preserving data into two independent modules that run Generate data and privacy tests. The data generation part needs to build a model, which can generate synthetic data according to the probability distribution of each attribute of the data and the statistical results of the data set, and ensure that its synthesis efficiency is high. This is a statistics-related task that requires a good understanding of the relevant characteristics of the original dataset. Relatively speaking, the purpose of the privacy test part is to ensure that the output data does not reveal sensitive privacy information. Every piece of synthetic data is subject to this privacy test — if the data passes it, it can be safely released, otherwise it is discarded. By adjusting the privacy test parameters, it can be observed that the published output data contains a certain amount of data that has the same impact on the final output statistical results. At this time, any output cannot allow an attacker to distinguish real data from synthetic data, achieving a "plausible" protection effect.

\subsection{A "denial" privacy protection mechanism}
For the generation model M and a specific data set D, a privacy protection mechanism F needs to be designed to ensure that all data finally released can meet the preset privacy standards. Specifically, Mechanism F builds a privacy-preserving mechanism in plausible denials with parameters by using synthetic data generated by model M on dataset D. In the following, the mechanism F to achieve privacy protection will be described:

For the given synthetic data, according to the parameters, judge whether the synthetic record y can be successfully output:

A piece of data d is randomly selected from the data set D, and a candidate synthetic record is generated according to the preset privacy parameters. Invoke the privacy test module according to the above parameters, if the set of parameters pass the privacy test, release the synthetic record y, otherwise discard it.
The core of Mechanism F is the parameter design of privacy testing. If a synthetic data record does not meet the privacy criteria for the given parameters, it is simply rejected. This paper argues that this criterion can be effectively enforced without measuring the sensitivity of the generative model M to the input data record D. It merely tests the environment, checking that the input data set contains enough plausible data records to reasonably judge whether to output synthetic data records. In order to judge whether the known parameters are processed and whether the piece of data can be released, it is necessary to make the data pass the deterministic test.

The privacy test T needs to meet a strict condition, that is, it is known that the probability of a certain seed d and the generated synthetic data y and the probability of another given data $d_a$ generating the same record are limited, where i is a positive integer. In this privacy test, different k values can be set to adjust the value range of the equation. In other words, increasing the value of k will increase the indistinguishability of released synthetic data y, and the number of trusted records that may produce y will also increase accordingly. Under this condition, an adversary with partial knowledge of the input data set cannot determine with high probability whether a particular input record d is is the seed for the output data y. However, whether y can pass the privacy test will reveal some information about the original seed size. This may lead the attacker to speculate whether a certain data d is contained in the input data set. In order to deal with this situation, the privacy parameter k can be randomized and transformed into a differential privacy protection mechanism controlled by parameters.

\subsection{Differential privacy}
In traditional data analysis, it is assumed that the data set available for query is extremely large, and the large data set only allows the query of information in the form of statistics. In theory, this method can protect all data from leakage. However, if the attacker knows some information about a specific piece of data contained in the database in advance (that is, the attacker has a knowledge background), he can use the differential attack method to obtain the detailed information of the data through conditional query. Even if supervision mechanisms such as prohibiting partial queries and anonymizing the database are introduced, privacy leaks cannot be avoided. The reason is that the decision to prohibit query itself will lead to privacy leakage; the attacker has a rich prior knowledge background, and can split a single destructive privacy query into a series of auxiliary information queries, and then pass the data Analysis can infer detailed private information with a high probability.

In response to the above problems, differential privacy needs to protect the published data from leaking sensitive information. Its definition can be expressed as: For the data contained in the data set, no matter whether any analysis is performed, or further merging with other information based on the analysis results Processing will not lead to the disclosure of personal privacy. By introducing randomness, such as differential privacy protection based on the Laplacian mechanism, the sensitivity and privacy budget are preset, and privacy protection is completed by adding white noise and sacrificing part of the usability.

\section{Data synthesis scheme design for privacy protection}
This section refers to some of the design ideas of Bindschaedler and others to design a privacy-preserving data synthesizer, focusing on the implementation process of the improved plausible denial mechanism and how to use this model to generate synthetic data. At the heart of this data synthesizer is a probability-based generative model that captures the joint distribution of attributes and then produces synthetic data based on the probabilities of the features output by the seeds in the real dataset. This requires extracting a sufficient number of training data samples from the selected real data set D for parameter learning after understanding the distribution characteristics of the data. In order to ensure the security of the published data, the model also needs to design the model of the privacy test part to protect the data samples so that their privacy will not be leaked. This paper will provide different privacy guarantees to achieve differential privacy protection with different degrees of strictness.

Before starting to work, assume that the data set D has three subsets DS, DT and DP with no intersection. These data sets will be applied to data synthesis and model structured and parameterized learning process respectively.

\subsection{Modelling}\label{AA}
In the following description, the joint probability distribution formula will be used to design the data synthesis model. The core of the model is to generate synthetic data based on real seeds based on the parameter distribution relationship between random variables and conditional probabilities captured by G.

The main functions of the data synthesizer can be summarized as follows: input a sensitive data set, and generate a synthetic data set according to the set parameters under the premise of ensuring that the sensitive privacy parameters are not leaked. Here, the data synthesis part is divided into three modules for implementation:

(1) Data conversion

This part is used to convert the input data set to obtain the joint probability distribution between the required eigenvalues. Therefore, it is necessary to calculate the statistical distance of the original records in this part. In addition, it is necessary to complete basic work in this part, such as setting the storage area of the software, setting interface functions, converting data formats, etc. In order to evaluate the efficiency of the model later, the time-consuming statistical output function is constructed in this part.

(2) Data generation

The generation module is the core part of the design, which is used to generate a series of candidate synthetic data according to the statistical characteristics of the input data. First, the processed data is read, and if its output is error-free, it is moved into the workspace and a path is generated. Afterwards, according to the privacy parameters entered in the parameter adjustment section, restart when the parameters are illegal, until a value that satisfies the set condition range is obtained. According to the entered parameters and attribute description documents, the processed data is resampled, and candidate synthetic data is generated according to the maximum scoring function.

(3) Data extraction

This module will integrate the generated products and collect and output the data that meets the privacy parameters, corresponding to the privacy test part of the mechanism. Specifically, in the code, firstly, according to the k value entered in the parameter setting, extract no less than this amount of candidate data, generate a composite storage file, and then call the privacy test for the product. A plausible denial mechanism is provided in the code, and differential privacy parameters are output, which can be measured to determine the privacy protection performance of the product.

\subsection{Data synthesis}
The data synthesis model constructed in this part needs to re-update its attributes according to the probability distribution of the data in the real data set, and then generate synthetic data y according to the probability distribution. According to the preset conditions in Section 3.1, it is known that ${x_1, x_2, ..., x_m}$ is taken from the data synthesis subset DS in the large data set D, which is the attribute set of the data records obtained by randomly sampling DS first. Distribution. On this basis, it is assumed that $\omega$ represents the numerical value of the newly generated attribute number. Then the data synthesizer will retain the original values of $m – \omega$ attributes when synthesizing data y (that is, copy the original values of this number of attributes from the seed of subset DS and assign them to y). In order to accurately define the resampling order of attributes, let the parameter $\sigma$ denote ${1, 2, ..., m}$, which is the sequence number for attribute reshaping.

\subsection{Structure learning}
The above paragraph describes the overall design principle of the data synthesizer and the specific functions that need to be realized. According to the probability distribution of the data in the data set DS, the data attributes are retained or updated according to the dependency order between the random variables mentioned above. In this paragraph, the design idea of structural learning will be adopted, and the synthetic data produced from the real data set DT and the seed data will be independent of each other (at least to achieve low correlation) through machine learning algorithms. These data can be captured by attackers with knowledge background.

The core of the algorithm is the optimal policy scoring (ie score maximization) function. This function is used to describe the correlation between a number of attributes and data, which has been described in many literatures. This article will use the feature (attribute) selection method to construct an algorithm model. At the general level, this method can be expressed as: from the many original characteristic attributes of the data itself, delete the attributes that have little impact on the output, so that the model based on a certain target structure can be simplified at the calculation level. In other words, this method can reduce the dimension of the data set and improve the efficiency of machine learning. In this paper, the algorithm design needs to find the best feature set among all the attributes of the data under the condition that the acyclic graph DAG remains acyclic, and add this set as the parent attribute for probability prediction.

Common feature selection based on filtering methods include variance selection method, correlation coefficient method, chi-square test method, etc., which can sort the original attributes of data according to the ability to predict predetermined feature values. In order to avoid ignoring the information redundancy between features as much as possible and improve the reliability of its feature subset in privacy protection, this paper adopts the feature selection method based on correlation coefficient to design part of the algorithm for structure learning. This choice can retain its versatility, fast training, low algorithm complexity and is suitable for large-scale data sets. The design flow chart is as follows:

In doing so, the algorithmic design of structural learning is transformed into a way to find optimal parameters, optimize the characteristics of the data, and use the products to select and target attributes. A qualified subset of optimized features needs to have features that are highly correlated with the target attribute. Furthermore, these features need to be independent of each other (at least with low correlation in practice). By optimizing to find the best feature subset, the complexity of the algorithm is reduced so that it can be applied to larger data sets.

From an overall point of view, the numerator is used as a reward fraction, the higher the correlation between the parent attribute and the target attribute, the higher the final score; the denominator punishes the correlation between the parent attributes, and when the attributes are closer to independence, the final score will be higher higher. Through the calculation of the scoring function, a higher score requirement is set as the judgment standard, and the subset that successfully passes the judgment will be output as the best subset, otherwise, repeat the steps of generating the subset until the program callable Memory cap or output best subset.

In this part, the final optimization goal of the construction algorithm is to maximize the total score PG(i) of all attributes i contained in the data set. It must be acknowledged, however, that the computational difficulty of searching for possible solutions to this optimal subset scales exponentially with the number of attributes contained in the data. Therefore, in the calculation of multiple eigenvalues of large data sets, the theoretical optimal solution cannot be found at this stage, and only approximate estimates can be made. According to literature review, a greedy algorithm can be used, that is, pre-assuming that the parent set of the target attribute is empty, and then continuously adding attributes to approximate the best score.

In order to get the best subset more quickly and accurately, in the current model design process, two constraints are added: first, it is necessary to consider that the DAG graph obtained from the original data feature set (ie, the parent set) should be acyclic, so that In order to satisfy the use of discrete conditional probability to calculate its degree of correlation and information entropy.

Secondly, it is necessary to pre-set the maximum complexity overhead for each attribute superset to prevent error reporting or unexpected results due to the program exceeding the memory range of the program. Since the computational cost is proportional to the number of possible joint distributions in the parent set, the computational cost will increase sharply as the number of attributes in the parent set increases.

In this formula, where $x_j$ is the total number of possible values that the attribute can take, this condition is added to prevent the appearance of too many supersets. As the number of feature attributes increases, the computational complexity increases exponentially, and the probability of finding the best subset in the same time will drop sharply. Adding this constraint will effectively improve the calculation amount of the algorithm and its memory usage. Although doing so may result in suboptimal subsets being obtained, this restriction effectively prevents overfitting of the conditional probabilities, causing the synthetic data to fall into low confidence intervals. From a generality perspective, algorithms that add constraints will perform more reliably most of the time.

In order to further save the calculation cost and improve the calculation speed, the algorithm model will discretize the continuous values of the attributes in the parent set, and the already discrete values will retain the original values. The idea is realized by designing a discretization function, dividing the data value into regions, and then updating its attribute value. The updated properties will be more self-contained. This approximation itself reduces the computational overhead of the superset while further preventing overfitting of the data.

At present, there is still a problem to be solved, that is, how the algorithm model can protect the privacy of the original data set. During the model design process, using the data in the dataset DT as a seed may lead to privacy leakage. The parameters required for structure learning used in this part can be transformed into the calculation of the occurrence probability of the data feature value and its joint probability distribution. Therefore, it is possible to add appropriate low-frequency white noise to the probability distribution of data to achieve differential privacy protection for structure learning. Due to the previous work, the correlation of features has been transformed to obtain the information entropy calculation of approximate random variables. Therefore, it is only necessary to keep the value range of the relevant measure unchanged, and then add white noise to change its entropy value.

\subsection{Privacy test parameter learning}
Continuing from the above, the article has already solved the problem of setting the framework of the data synthesis model, and transformed the mathematical problem into a statistical problem. In this paragraph, in order to obtain privacy test parameters that can satisfy differential privacy, it is necessary to perform parameter learning on the data set DP to obtain its conditional probability. In order to perform parameter learning, it is first necessary to obtain the prior distribution of the conditional probability parameters, and calculate the learning hyperparameters (that is, the prior distribution parameters of the model) from the DP according to the prior distribution, so as to infer the overall conditional probability of the data set D.

Here, the data used in the calculation is DP, which is a discrete attribute distribution. For the convenience of modeling, it is assumed that there is a descriptive distribution polynomial among the eigenvalues, and the data are grouped. It is further assumed that the conjugate prior distribution of the polynomial satisfies the Dirichlet condition. Therefore, the Dirichlet distribution can be used to obtain statistical information from a set of data and set distribution probabilities for most polynomials.

The above calculations all need to satisfy a premise: all attributes, according to their occurrence times, can be transformed into discrete probability distributions. If one or several attributes have continuous values, data preprocessing is required, and the regression function included in the machine learning library is called to discretize its values. This part is relatively simple, so it is not repeated. After this operation, a random subset generated from the original dataset will output an approximate distribution.

It should also be noted that since the specific parameters of the conditional probability depend on the data records in the DP, publishing them leads to privacy security issues. Attackers may use differential attacks, which will increase the risk of some sensitive information being leaked. Therefore, before publishing conditional probability parameters, it is necessary to make the parameters meet the relevant conditions of differential privacy guarantee. In the parameter learning part, DP is only used to calculate $n_i^C$, a vector describing the data in the parent set. In order to calculate the variance of the power of adding white noise, and to achieve differential privacy protection, it is necessary to calculate the sensitivity of $n_i^C$ to each data record.

\section*{Conclusion}

Based on the existing differential privacy mechanism model, this paper designs a new plausible denial mechanism, and designs a data synthesis model based on it. This model can efficiently synthesize data, effectively solve the problem of difficult and time-consuming calculation of high-dimensional data sets under the differential privacy mechanism, and the quality of generated data is only related to the performance of the synthesizer. Using virtual machine verification, the data synthesis efficiency of the model is significantly better than that of common reinforcement learning models, and it is suitable for data sets with higher data dimensions, and has universal applicability. 

This verification classifier correct rate only adopts simple binary classification, and it is impossible to determine the specific performance of the model when using multi-class data. You can further use other label tests of this data set; the classifier used is not the best in classification accuracy The classifier should be tested with a higher algorithm model (such as random forest classifier, neural network classifier) that can distinguish accurately; new evaluation indicators, such as robustness, can be added to describe data synthesis more accurately The utility of privacy protection of the model; the simulation program is further expanded and packaged as software to realize privacy protection.

\bibliographystyle{unsrt}
\bibliography{conference_101719}

\end{document}